 \documentclass[final,5p,times,twocolumn]{elsarticle}

\usepackage{amssymb}
\usepackage{amsmath}
\usepackage{bm}

\journal{Journal of Subatomic Particles and Cosmology}

\begin{document}

\begin{frontmatter}



\title{Symmetric tensor portals to dark matter I: Creation of dark scalars}


\author{Brian Dignean}
\ead{brian.dignean@usask.ca}

\author{Alexander J.~Magnus}
\ead{alexander.magnus@usask.ca}

\author{Rainer Dick\corref{cor1}}
\ead{rainer.dick@usask.ca}

\affiliation{organization={Department of Physics and Engineering Physics, 
University of Saskatchewan},
            addressline={116 Science Place}, 
            city={Saskatoon},
            postcode={S7N 5E2}, 
            state={SK},
            country={Canada}}

\begin{abstract}
  We examine the creation of scalar dark matter through a symmetric tensor portal.
  We find that both freeze-in and thermal freeze-out through a symmetric tensor mediator can create scalar dark matter. The required tensor masses for freeze-in
  are much larger than for freeze-out for the same dark-matter mass, in agreement with the nonthermalization assumption in freeze-in scenarios.
\end{abstract}




\end{frontmatter}



\section{Introduction\label{sec:intro}}

Massive tensor fields are a hallmark of many attempts to formulate theories
of quantum gravity. For example, string theory predicts towers of massive symmetric
and antisymmetric tensor fields, both from string excitations
and through Kaluza-Klein modes, in addition to massless
tensor fields in the closed-string sector.
Although the towers of massive string states had been considered far out of reach
in traditional string theory, both the prospects of low-scale string models
\cite{anton1,cvetic,anchor1,ibanez1,ibanez2,karozas}
and of large extra dimensions \cite{anchor2,vafa}
have made them interesting targets for particle phenomenology beyond the Standard Model.
Antisymmetric tensors can also mediate
gauge interactions between strings \cite{KR}, and this has motivated recent studies
of antisymmetric tensors as means of inducing a $U_Y(1)$ dipole portal
to dark matter \cite{adrd,rdick},
antisymmetric tensor dark matter from freeze-in \cite{evan},
and antisymmetric tensor portals to dark matter \cite{amjfrd,plantier}.

We also note that massive
symmetric tensors are not just a feature of string excitations, but also figure
prominently in the recent investigations of massive gravity theories as alternatives
to standard Einstein gravity
\cite{horava,deRham1,hassan1,hassan2,hinterbichler1,alberte,deRham2,joyce,hinterbichler2}.
Indeed, these
theories have also spurred investigations of massive gravity for gravitational
waves \cite{elhadj}, compact astrophysical objects \cite{shahzad,bhar,chaudhary,yerra},
and cosmology \cite{kenjale,kazempour}.

The studies of antisymmetric and symmetric tensor portals are complementary.
Antisymmetric tensors couple in leading order to electric or magnetic dipole moments
of fermions \cite{adrd,rdick}, and the resulting contributions to Bhabha and M\"oller
scattering provide limits on tensor masses and couplings \cite{strd,malta} (and also
interesting search opportunities for antisymmetric tensors).

Symmetric tensors, on the other hand, couple both to bosons and to fermions, and
therefore they provide natural portals to both bosonic and fermionic dark matter.
In the present paper, we discuss the production of scalar dark matter.
We introduce the relevant couplings of a massive symmetric tensor
and summarize the relevant decay rates and annihilation
cross sections in Sec.~\ref{sec:couple}. Mass-coupling relations for dark scalars from
freeze-in through the symmetric tensor portal
will be reported in Sec.~\ref{sec:freezein}.
The corresponding thermal freeze-out production is discussed in Sec.~\ref{sec:freezeout}.
Sec.~\ref{sec:conc} summarizes our conclusions. We focus our calculations on the tens of TeV
mass ranges, because these are the mass ranges for beyond-the-Standard-Model physics
that we may hope to explore with next-generation colliders and direct-search experiments.

\section{Couplings for a symmetric tensor portal\label{sec:couple}}

We are interested in the implications of a massive symmetric traceless tensor $s_{\mu\nu}=s_{\nu\mu}$,
$s_\mu{}^\mu=0$. The primary motivation for this is the emergence of massive counterparts
of graviton states in string theory and in theories of massive gravity. From a particle
physics perspective, the trace part of any massive graviton-like tensor would constitute
an extra scalar degree of freedom. However, 
if the trace part $h$ of the massive symmetric tensor is much heavier than the traceless part,
$m_h\gg m_s$, the canonical interaction picture mode expansion for the traceless
part can be written for $E\sim m_s\ll m_h$ as
\begin{eqnarray}\nonumber
  s^{\mu\nu}(x)&\!\!\!\!=&\!\!\!\!\int\!\frac{d^3\bm{k}}{\sqrt{(2\pi)^32E(\bm{k})}}\,
  \sum_{\alpha,\beta}\epsilon_{\alpha}^\mu(\bm{k})\epsilon_\beta^\nu(\bm{k})
  \\ \label{eq:mode1}
  &&\!\!\!\!\times\left[
    a_{\alpha\beta}(\bm{k})\exp(\mathrm{i}k\cdot x)
    +a^+_{\alpha\beta}(\bm{k})\exp(-\,\mathrm{i}k\cdot x)\right],
\end{eqnarray}
with annihilation operators
\begin{equation}
  a_{\alpha\beta}(\bm{k})=a_{\beta\alpha}(\bm{k}),\quad
  \sum_{\alpha=1}^3 a_{\alpha\alpha}(\bm{k})=0,
\end{equation}
and canonical commutation relations
\begin{equation}
  [a_{\alpha\beta}(\bm{k}), a^+_{\gamma\delta}(\bm{k}')]=\frac{1}{2}
  \left(\delta_{\alpha\gamma}\delta_{\beta\delta}+\delta_{\alpha\delta}\delta_{\beta\gamma}\right)
  \delta(\bm{k}-\bm{k}').
  \end{equation}
The equations of motion for $s_{\mu\nu}$, which lead to the mode expansion (\ref{eq:mode1}),
are explained in the Appendix.

The polarization vectors in (\ref{eq:mode1}) are the standard 
vectors $\epsilon_\alpha^\mu(\bm{k})$, such that
$\epsilon_1^\mu(\bm{k})$ and $\epsilon_2^\mu(\bm{k})$
are spatial orthonormal vectors without time-like components and 
perpendicular to $\bm{k}$,
\begin{equation}\label{eq:sumalphaLG}
\sum_{\alpha=1}^2\bm{\epsilon}_\alpha(\bm{k})\otimes
\bm{\epsilon}_\alpha(\bm{k})=\underline{1}-\hat{\bm{k}}
\otimes\hat{\bm{k}},
\end{equation}
while the third polarization vector is
\begin{equation}\label{eq:epsilon3}
\epsilon_3(\bm{k})=\frac{(|\bm{k}|,k^0\hat{\bm{k}})}{\sqrt{
-\,k^2+\mathrm{i}\epsilon}}.
\end{equation}
The polarization vectors yield projectors
\begin{equation}
  P_\perp^{\mu\nu}(\bm{k})=\sum_{\alpha=1}^3\epsilon^\mu_\alpha(\bm{k})
  \epsilon^\nu_\alpha(\bm{k})=\eta^{\mu\nu}-\frac{k^\mu k^\nu}{k^2-\mathrm{i}\epsilon},
\end{equation}
where on-shell $k^2+m_s^2=0$.

Since our primary motivation for this study are massive partners of gravitons
through towers of massive states in low-scale string theory or large-volume compactifications,
or models of massive gravity,
our couplings to baryons and dark matter are informed by the leading order graviton couplings.

We note that all matter fields couple to $\sqrt{-\,g}\simeq 1+h$, where we
used the weak-field expansion (\ref{eq:weakfield1}). This factor does not concern us
in cases where the massive graviton-like fields arise from a tower of massive states above
a massless graviton, when $h$ would be the trace part in the massless sector.
Furthermore, the factor also does not concern us if the symmetric tensor portal arises from
massive gravity, since we anyway assume $m_h\gg m_s$. For this reason, we will 
only consider couplings of the symmetric traceless tensor $s_{\mu\nu}$ in the following.


The coupling of $s_{\mu\nu}$ to real scalar fields $\phi$ follows from the kinetic term
in the form
\begin{equation}\label{eq:couplephi}
\mathcal{L}_\psi=\frac{1}{f_s}s^{\mu\nu}\partial_\mu\phi\cdot\partial_\nu\phi.
\end{equation}
Here, we absorbed the mass scale $f_s=\kappa^{-1/2}$ into the tensor field for
canonical mass dimension 1, see the Appendix.

The coupling of $s_{\mu\nu}$ to gauge fields $V_\mu^I$ with field strengths
$V_{\mu\nu}^K= \partial_\mu V_\nu^K-\partial_\nu V_\mu^K+gV_\mu^I V_\nu^Jf_{IJ}{}^K$,
also follows from their kinetic terms as
\begin{equation}\label{eq:coupleV}
\mathcal{L}_V=\frac{1}{f_s}s^{\mu\nu}V_{\mu\rho}^K V_{K\nu}{}^{\rho}.
\end{equation}

For the coupling to Dirac fields, we note that we can identify
\begin{equation}
  h_{\mu\nu}=s_{\mu\nu}+\frac{1}{4}\eta_{\mu\nu}h
\end{equation}
with a weak-field expansion of the tetrad, $e_\mu{}^a=\eta_\mu{}^a+\epsilon_\mu{}^a$,
through $h_{\mu\nu}=\epsilon_{\mu\nu}$, by simply choosing a symmetric solution to
the tetrad condition $g_{\mu\nu}=e_\mu{}^a e_\nu{}^b\eta_{ab}$ in the weak-field limit.
This yields a coupling of $s_{\mu\nu}$ to Dirac fields through the appearance
of the tetrad in the kinetic term,
\begin{equation}
  \mathcal{L}_\psi=\frac{\mathrm{i}}{2}
  e^\mu{}_a\left(\overline{\psi}\gamma^a\partial_\mu\psi
  -\partial_\mu\overline{\psi}\cdot\gamma^a\psi\right),
\end{equation}
where the first term contributes through
\begin{equation}\label{eq:spsi0}
  e^\mu{}_a\overline{\psi}\gamma^a\partial_\mu\psi
  =\overline{\psi}\gamma^\mu\partial_\mu\psi
  +s^{\mu\nu}\overline{\psi}\gamma_\nu\partial_\mu\psi,
\end{equation}
and similarly for the second term.
As before, trace terms $h$ have been omitted in Eq.~(\ref{eq:spsi0}).
The corresponding coupling terms after absorbing the mass scale $f_s$ are then 
\begin{equation}\label{eq:spsi}
  \mathcal{L}_\psi=\frac{\mathrm{i}}{2f_s}s^{\mu\nu}
  \left(\overline{\psi}\gamma_\nu\partial_\mu\psi-\partial_\mu\overline{\psi}\cdot\gamma_\nu\psi
  \right).
\end{equation}
  
Furthermore, one might suspect that a coupling to spin connections might also
contribute if tetrad fields are used to gauge the Lorentz group.
Gauging the Lorentz group involves inclusion of a term
\begin{equation}
  \mathcal{L}_\Omega=\frac{\mathrm{i}}{2}\overline{\psi}
  \left(\gamma^\mu\Omega_\mu+\Omega_\mu\gamma^\mu\right)\psi
\end{equation}
in the Dirac Lagrangian, where
\begin{equation}
  \Omega_\mu=\frac{\mathrm{i}}{2}\Gamma_{ab\mu}S^{ab}
  =-\,\frac{1}{4}\Gamma_{ab\mu}\gamma^a\gamma^b
\end{equation}
is the spin connection involving the Dirac representation of
Lorentz generators, $S^{ab}=\mathrm{i}[\gamma^a,\gamma^b]/4$.
However, this will not contribute a leading order coupling
to $s_{\mu\nu}$ by virtue of
\begin{eqnarray}\nonumber
  \left.\{\gamma^\mu,\Omega_\mu\}\right|_{\mathcal{O}(s)}&\!\!\!\!=&\!\!\!\!
 \frac{1}{4}\left(\gamma^\mu\gamma^b\gamma^c
  +\gamma^b\gamma^c\gamma^\mu
  \right)
  \left(\partial_b s_{c\mu}-\partial_c s_{b\mu}\right)
  \\ \nonumber
   &\!\!\!\!=&\!\!\!\!\frac{1}{4}\left(\gamma^\mu\gamma^b\gamma^c\partial_b s_{c\mu}
  -\gamma^b\gamma^c\gamma^\mu\partial_c s_{b\mu}
  \right)
  \\
  &\!\!\!\!=&\!\!\!\!
\frac{1}{4}\left(\gamma^\mu\gamma^b\gamma^c
  -\gamma^c\gamma^b\gamma^\mu
  \right)\psi\partial_b s_{c\mu}=0.
\end{eqnarray}
where we used $\gamma^c\gamma^\mu s_{c\mu}=-\,s^\mu{}_{\mu}=0$
in the second line and the symmetry of $s_{c\mu}$ in the final step.

We need the decay constant $\Gamma_s$ of the tensor in the calculation
of the dark matter production cross sections through the symmetric tensor portal.
The coupling (\ref{eq:couplephi}) yields the decay widths into scalars
with masses $m_\phi<m_s/2$
\begin{equation}
  \Gamma_{s\to\phi\phi}
  =\frac{1}{1152\pi f_s^2 m_s^2}(m_s^2-4m_\phi^2)^{5/2}.
\end{equation}
The coupling (\ref{eq:coupleV}) yields the decay widths into the
gauge bosons with masses $m_V<m_s/2$,
\begin{equation}
  \Gamma_{s\to\gamma\gamma}=\frac{7m_s^3}{2304\pi f_s^2},
\end{equation}
\begin{equation}\label{eq:Gamma2gg}
  \Gamma_{s\to gg}=\frac{7m_s^3}{288\pi f_s^2},
\end{equation}
\begin{equation}\label{eq:Gamma2ZZ}
  \Gamma_{s\to ZZ}=
  \left.\frac{A_Z(p)\sqrt{m_s^2-4m_Z^2}}{144\pi f_s^2 m_s^2}\right|_{p=\sqrt{m_s^2-4m_Z^2}\big/2},
\end{equation}
\begin{equation}\label{eq:Gamma2WW}
  \Gamma_{s\to W^+W^-}=
  \left.\frac{A_W(p)\sqrt{m_s^2-4m_W^2}}{72\pi f_s^2 m_s^2}\right|_{p=\sqrt{m_s^2-4m_W^2}\big/2}.
\end{equation}
The momentum-dependent factors in the decay constants (\ref{eq:Gamma2ZZ},\ref{eq:Gamma2WW})
are
\begin{equation}
  A_V(p)=7p^4+7p^2 m_V^2+3m_V^4.
\end{equation}
Finally, the coupling (\ref{eq:spsi}) yields the decay width into fermion
pairs $f\overline{f}$,
\begin{equation}\label{eq:s2ff}
  \Gamma_{s\to f\overline{f}}=\frac{N_c}{144\pi f_s^2 m_s^2}(m_s^2-4m_f^2)^{3/2}
  (3m_s^2+8m_f^2).
\end{equation}
The color factor $N_c$ in Eq.~(\ref{eq:s2ff}) is $N_c=1$ for leptons and $N_c=3$ for quarks.

The calculations of scalar dark matter creation through freeze-in or thermal freeze-out
through a symmetric tensor portal also require the corresponding annihilation cross sections
for annihilation of baryons to dark scalars. We denote the mass of the real dark scalar $D$
as $m_d$.

\begin{figure}[thb]
\scalebox{0.55}{\includegraphics{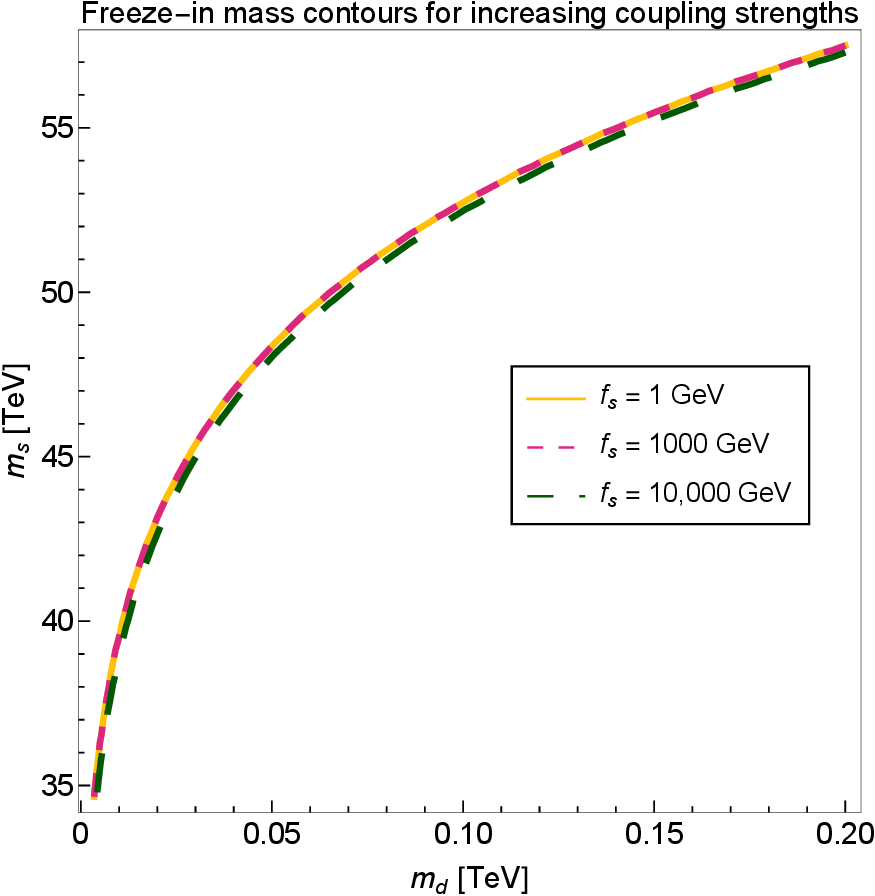}}
\caption{\label{fig:FIr1}
Relation between dark-matter mass and tensor mass for dark-matter masses in the 100 GeV region.}
\end{figure}

The couplings (\ref{eq:coupleV}-\ref{eq:spsi}) yield the baryon
annihilation cross sections for $\sqrt{s}>2m_d$,
\begin{eqnarray}\label{eq:hh2DD}
  v\sigma_{hh\to DD}&\!\!\!\!=&\!\!\!\!\frac{(s-4m_d^2)^{5/2}}{1280\pi f_s^4s^{3/2}}\frac{
    (s-4m_h^2)^2}{(s-m_s^2)^2+m_s^2\Gamma_s^2},
  \\ \label{eq:ff2DD}
  v\sigma_{f\overline{f}\to DD}&\!\!\!\!=&\!\!\!\!
  \frac{(s-4m_d^2)^{5/2}}{3840\pi f_s^4s^{3/2}}
  \frac{(s-4m_f^2)(s+6m_f^2)}{(s-m_s^2)^2+m_s^2\Gamma_s^2},
  \\ \nonumber
  v\sigma_{\gamma\gamma\to DD}&\!\!\!\!=&\!\!\!\!v\sigma_{gg\to DD}
  \\ \label{eq:gammagamma2DD}
  &\!\!\!\!=&\!\!\!\!\frac{23\sqrt{s}}{7680\pi f_s^4}
  \frac{(s-4m_d^2)^{5/2}}{(s-m_s^2)^2+m_s^2\Gamma_s^2},
  \\ \label{eq:ZZ2DD}
  v\sigma_{ZZ\to DD}&\!\!\!\!=&\!\!\!\!\frac{(s-4m_d^2)^{5/2}}{17280\pi f_s^4s^{3/2}}
   \frac{B_Z(s)}{(s-m_s^2)^2+m_s^2\Gamma_s^2},
 \\ \label{eq:WW2DD} 
 v\sigma_{W^+W^-\to DD}&\!\!\!\!=&\!\!\!\!\frac{(s-4m_d^2)^{5/2}}{17280\pi f_s^4s^{3/2}}
 \frac{B_W(s)}{(s-m_s^2)^2+m_s^2\Gamma_s^2},
\end{eqnarray}
for the forward balance equations in the freeze-in scenario.
The factors $B_V(s)$ are
\begin{equation}
B_V(s)=23s^2-104m_V^2 s+168m_V^4.
\end{equation}

\begin{figure}[thb]
\scalebox{0.55}{\includegraphics{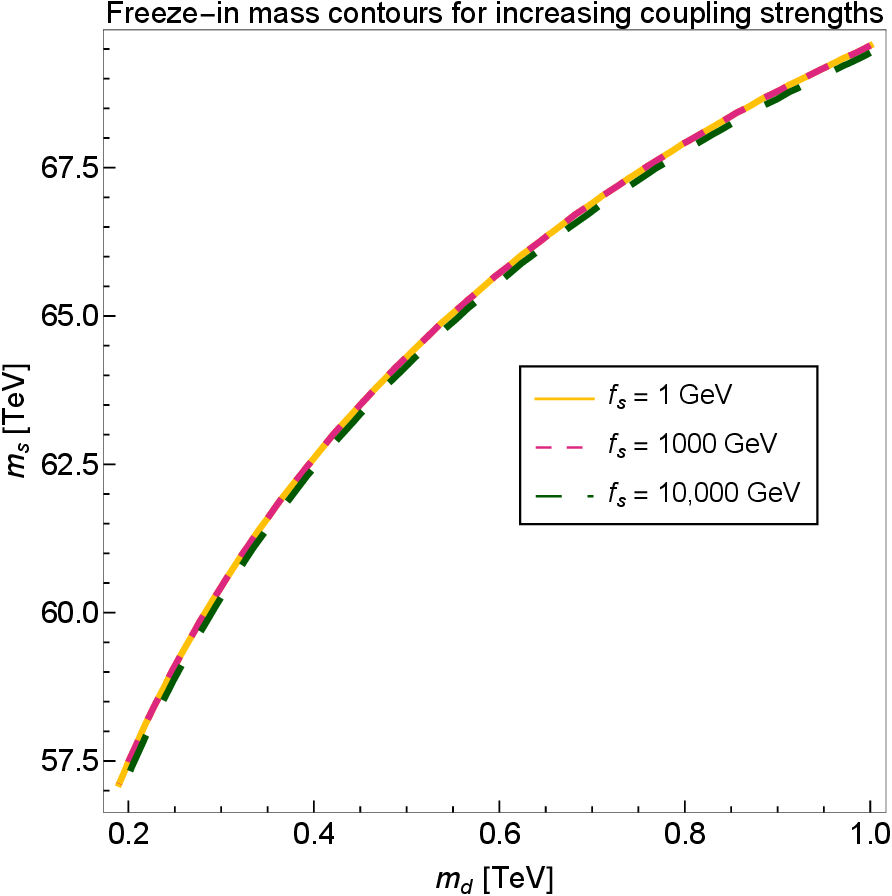}}
\caption{\label{fig:FIr2}
Relation between dark-matter mass and tensor mass for dark-matter masses between 200 GeV and 1 TeV.}
\end{figure}

Thermal freeze-out involves principally both forward and backward cross sections.
However, the balance equation for thermal freeze-out
can conveniently be expressed using only the
backward cross sections \cite{KT}. The couplings (\ref{eq:coupleV}-\ref{eq:spsi}) yield
the following dark-matter annihilation cross sections through a massive tensor portal,
\begin{eqnarray}\label{eq:DD2hh}
  v\sigma_{DD\to hh}&\!\!\!\!=&\!\!\!\!\frac{(s-4m_h^2)^{5/2}}{1280\pi f_s^4s^{3/2}}\frac{
    (s-4m_d^2)^2}{(s-m_s^2)^2+m_s^2\Gamma_s^2},
  \\
  \label{eq:DD2ff}
  v\sigma_{DD\to f\overline{f}}&\!\!\!\!=&\!\!\!\!N_c
  \frac{(s-4m_d^2)^2}{480\pi f_s^4s^{3/2}}
  \frac{(s-4m_f^2)^{3/2}(s+6m_f^2)}{(s-m_s^2)^2+m_s^2\Gamma_s^2},
  \\ \nonumber
  v\sigma_{DD\to\gamma\gamma}&\!\!\!\!=&\!\!\!\!v\sigma_{DD\to gg}
  \\ \label{eq:DD2gammagamma}
  &\!\!\!\!=&\!\!\!\!\frac{23s}{1920\pi f_s^4}
  \frac{(s-4m_d^2)^2}{(s-m_s^2)^2+m_s^2\Gamma_s^2},
 \\ \label{eq:DD2ZZ}
 v\sigma_{DD\to ZZ}&\!\!\!\!=&\!\!\!\!\frac{(s-4m_d^2)^2}{1920\pi f_s^4s^{3/2}}
 \frac{B_Z(s)(s-4m_Z^2)^{1/2}}{(s-m_s^2)^2+m_s^2\Gamma_s^2},
 \\ \label{eq:DD2WW}
 v\sigma_{DD\to W^+W^-}&\!\!\!\!=&\!\!\!\!\frac{(s-4m_d^2)^2}{960\pi f_s^4s^{3/2}}
  \frac{B_W(s)(s-4m_W^2)^{1/2}}{(s-m_s^2)^2+m_s^2\Gamma_s^2},
\end{eqnarray}
which are directly related to (\ref{eq:hh2DD}-\ref{eq:WW2DD}) through the fact that
the cross sections involve the momentum $p=\sqrt{s-4m^2}/2$ of the produced particles,
cross sections into Majorana particles involve an extra factor $1/2$ from the degeneracy of
the final state, and cross sections are averaged over initial polarizations, but summed over
final polarizations.

The color factor $N_c$ in Eq.~(\ref{eq:DD2ff})
appears in $v\sigma_{DD\to f\overline{f}}$ due to summation over possible final states, whereas
$v\sigma_{f\overline{f}\to DD}$ involves averaging over initial states, similar to spin
polarizations.

\section{Freeze-in dark matter through the
  symmetric tensor portal\label{sec:freezein}}

\begin{figure}[htb]
\scalebox{0.55}{\includegraphics{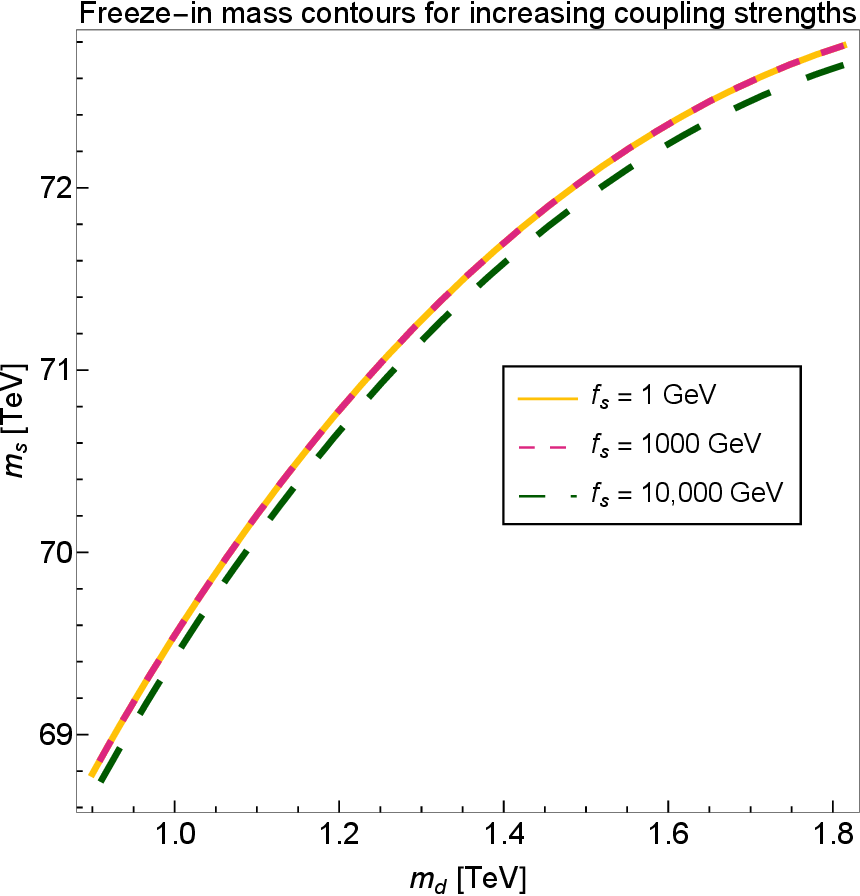}}
\caption{\label{fig:FIr3}
Relation between dark-matter mass and tensor mass for dark-matter masses between 1 TeV and 1.8 TeV.}
\end{figure}

 The creation of dark matter from freeze-in involves forward integration of the baryon-to-DM
 annihilation equations \cite{mcdonald2002,jmr2010}.
Here, this involves the baryon-to-DM annihilation cross sections (\ref{eq:hh2DD}-\ref{eq:WW2DD}),
which are independent of the coupling scale $f_s$ near the resonance $s=m_s^2$.
We find that integration of the baryon-to-DM creation equations for freeze-in of
dark scalars through a symmetric tensor portal
reveals very little $f_s$-dependence of the relation between symmetric tensor mass $m_s$
and dark matter mass $m_d$, see Figs.~\ref{fig:FIr1}-\ref{fig:FIr3}. This shows that integration of
the dark matter creation equations is dominated by the pole region $s\simeq m_s^2$.

\begin{figure}[htb]
\scalebox{0.55}{\includegraphics{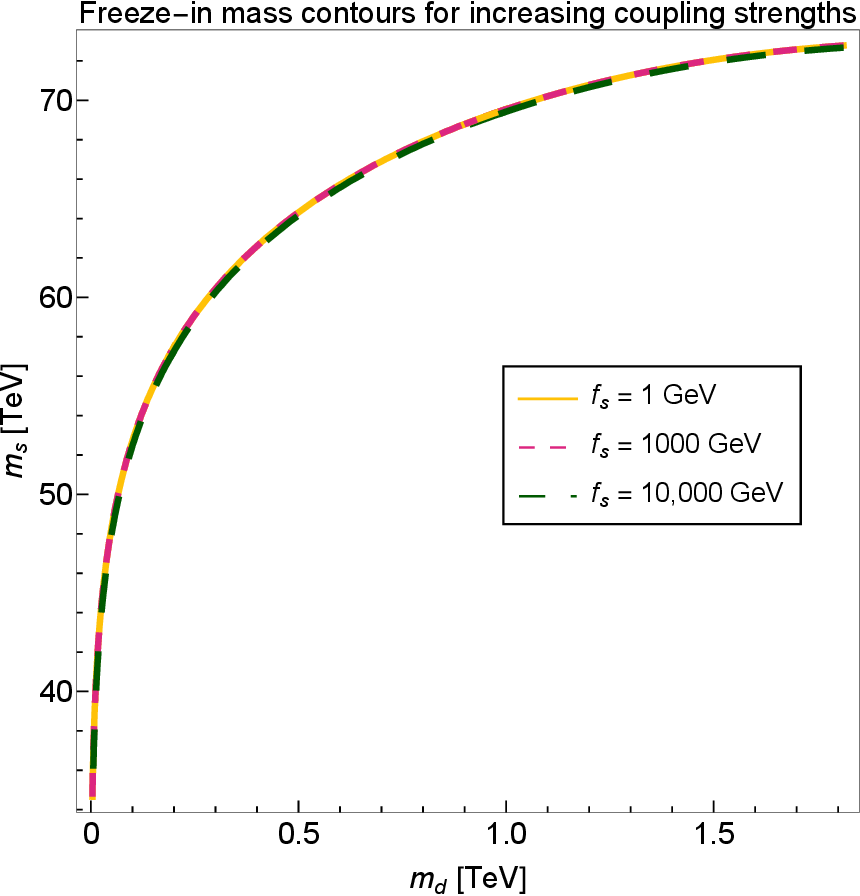}}
\caption{\label{fig:freezein}
Relation between dark-matter mass and tensor mass for dark-matter masses up to 1.8 TeV.}
\end{figure}

We note that weaker coupling (i.e.~increasing $f_s$) gradually increases the dark matter
mass $m_d$ for given tensor mass $m_s$. 
This stems from the fact that increasing $f_s$ yields
slower freeze-in which leads to fewer relic dark-matter particles. Matching the observed 
dark-matter mass density therefore requires higher dark-matter mass.

\section{Freeze-out dark matter through the
  symmetric tensor portal\label{sec:freezeout}}

\begin{figure}[htb]
\scalebox{0.55}{\includegraphics{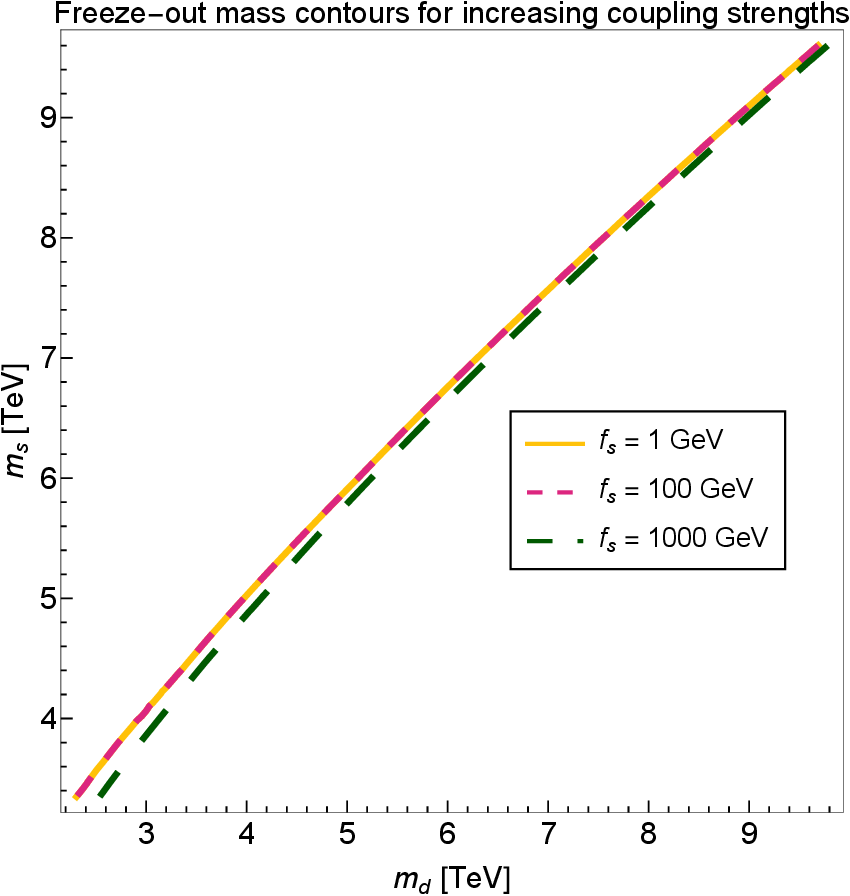}}
\caption{\label{fig:FOr2}
Relation between dark-matter mass and tensor mass for dark-matter masses between 2 TeV and 10 TeV.}
\end{figure}

\begin{figure}[htb]
\scalebox{0.55}{\includegraphics{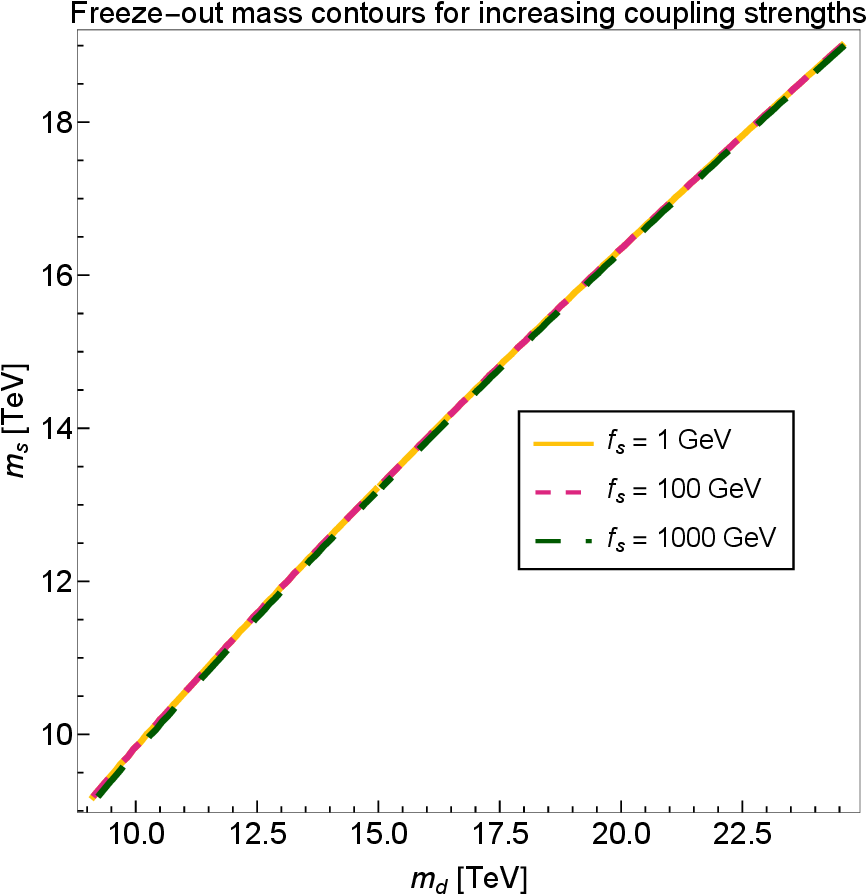}}
\caption{\label{fig:FOr3}
Relation between dark-matter mass and tensor mass for dark-matter masses between 9 TeV and 24.5 TeV.}
\end{figure}

\begin{figure}[htb]
\scalebox{0.55}{\includegraphics{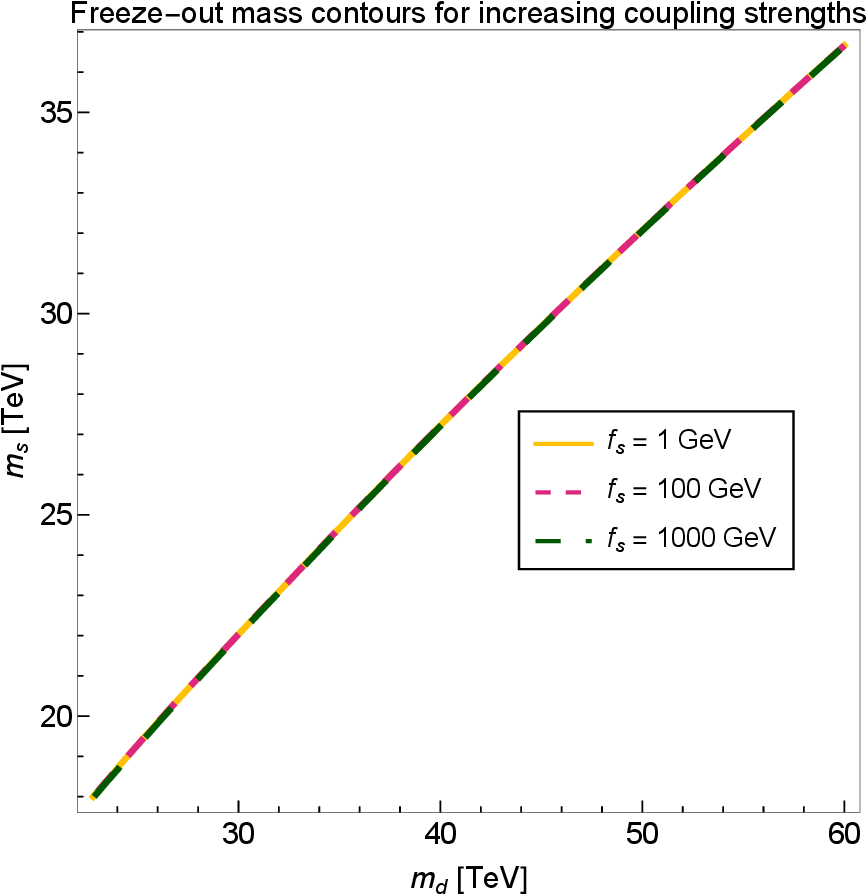}}
\caption{\label{fig:FOr4}
Relation between dark-matter mass and tensor mass for dark-matter masses between 24 TeV and 60 TeV.}
\end{figure}

We also find that freeze-out of dark matter through the symmetric
tensor portal is dominated by the pole region $s\simeq m_s^2$,
which again leads to very weak dependence on $f_s$, as shown 
in Figs.~\ref{fig:FOr2}-\ref{fig:FOr4}.

We find again that increasing $f_s$
requires larger dark matter mass for given tensor mediator mass. 
However, we could not find solutions for the dark matter abundance conditions anymore
for very large $f_s$ in the freeze-out scenario. The corresponding very weak couplings
apparently do not comply with the requirement of initial thermalization with the baryonic
heat bath in the freeze-out scenario.

\begin{figure}[htb]
\scalebox{0.55}{\includegraphics{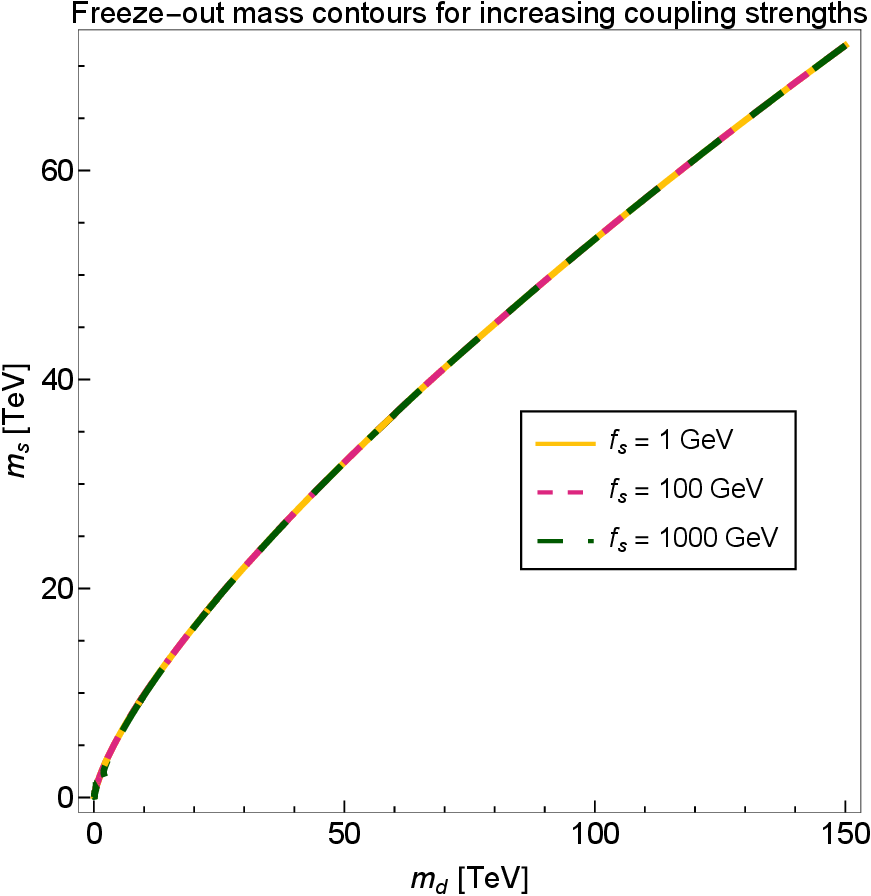}}
\caption{\label{fig:freezeout}
Relation between dark-matter mass and tensor mass for dark-matter masses up to 150 TeV.}
\end{figure}

We also note that, for the same dark matter mass $m_d$ and coupling $f_s$, freeze-in 
requires much larger mediator masses than freeze-out.
This complies with the assumption that dark matter never reached thermal equilibrium with
the baryonic heat bath in the freeze-in scenario, because larger mediator mass implies weaker 
coupling between the baryonic and dark sectors.

\section{Conclusions\label{sec:conc}}

We have found relations between symmetric tensor masses and scalar dark matter masses under the assumption
that the dark matter is either created through freeze-in (Fig.~\ref{fig:freezein})
or thermal freeze-out from the primordial heat bath (Fig.~\ref{fig:freezeout}). We have focused on 
mass values up to about 100 TeV. The results for freeze-in yield $m_d\ll m_s$, whereas 
thermal freeze-out yields $m_d\lesssim m_s$ in the TeV mass range and $m_d\gtrsim m_s$ in the 
tens-of-TeV mass range. The observation of much larger mediator masses for given parameters $m_d$
and $f_s$ in the freeze-in scenario is compatible with the assumption of absence of
thermalization between baryons and dark matter in that scenario, because larger mediator mass implies
weaker effective coupling between the baryonic and dark sectors.
Weaker coupling in the freeze-in scenario was also observed for antisymmetric tensor
mediators \cite{amjfrd}, and this is in general a consistency condition for any 
portal that might produce dark matter either through freeze-in or thermal freeze-out.
This implies that freeze-in is inherently more efficient than thermal freeze-out.
Apparently, the stronger couplings in the thermal freeze-out scenarios would then overproduce dark
matter if the dark matter particles in those scenarios would not exist in thermal equilibrium
with the primordial heat bath between the epochs of reheating and thermal freeze-out.

\appendix

\section{The traceless symmetric tensor field in the interaction picture}

Our motivation to study symmetric tensor fields arises from massive partners
of gravitons or from massive gravity.
Therefore, we recall the graviton expansion of the Einstein-Hilbert action.
Expansion of $\mathcal{L}_E=\sqrt{-\,g}R/2\kappa$ for a metric
\begin{equation}\label{eq:weakfield1}
g_{\mu\nu}=\eta_{\mu\nu}+2h_{\mu\nu}=\eta_{\mu\nu}+2s_{\mu\nu}+\frac{1}{2}\eta_{\mu\nu}h,
\end{equation}
with $h=\eta^{\mu\nu}h_{\mu\nu}$,
yields up to total derivative terms the second order expansion
\begin{eqnarray}\nonumber
  \kappa\mathcal{L}_E&\!\!\!\!=&\!\!\!\!\frac{1}{2}\partial_\mu s_{\nu\lambda}
  \cdot(\partial^\nu s^{\mu\lambda}+\partial^\lambda s^{\mu\nu})
  -\frac{1}{2}\partial_\mu s_{\nu\lambda}\cdot\partial^\mu s^{\nu\lambda}
  \\
  &&-\,\frac{1}{2}\partial_\mu s^{\mu\nu}\cdot\partial_\nu h
  +\frac{3}{16}\partial_\mu h\cdot\partial^\mu h,
\end{eqnarray}
where all indices are pulled with the Minkowski metric.

We also add a mass term
\begin{equation}
\kappa\mathcal{L}_m=\frac{m_h^2}{8}h^2-\frac{m_s^2}{2}s_{\nu\lambda}s^{\nu\lambda}.
\end{equation}
Taking a particle physics perspective requires independent variation
of $\mathcal{L}=\mathcal{L}_E+\mathcal{L}_m$ with
respect to $s_{\mu\nu}$ and $h$. This yields equations of motion
\begin{equation}\label{eq:s1a}
  \partial^2 s^{\nu\lambda}-\partial_\mu(\partial^\nu s^{\mu\lambda}+\partial^\lambda
  s^{\mu\nu})+\frac{1}{2}\partial^\nu\partial^\lambda h
  =m_s^2 s^{\nu\lambda}
\end{equation}
and
\begin{equation}\label{eq:h}
\partial^2 h=m_h^2 h,
\end{equation}
where the trace of Eq.~(\ref{eq:s1a}),
\begin{equation}\label{eq:s1b}
\partial_\mu\partial_\nu s^{\mu\nu}=\frac{1}{4}\partial^2 h,
\end{equation}
has been used to simplify Eq.~(\ref{eq:h}).

Imposing harmonic (Lorentz, de Donder) gauge
\begin{equation}
  \partial_\mu s^{\mu\nu}=\frac{1}{4}\partial^\nu h,
\end{equation}
decouples $h$ from Eq.~(\ref{eq:s1a}),
\begin{equation}\label{eq:s2}
  \partial^2 s^{\nu\lambda}=m_s^2 s^{\nu\lambda}.
\end{equation}

However, we cannot impose harmonic gauge, nor do 
we need it in the present investigation. Since we are interested in the
possible low-energy implications of massive symmetric tensors, not of
additional massive scalar fields, we assume $m_h\gg m_s$, such that
Eq.~(\ref{eq:s1b}) implies for energies well below $m_h$ that
$\partial_\mu\partial_\nu s^{\mu\nu}=0$, and Eq.~(\ref{eq:s1a}) then
also yields through contraction with $\partial_\nu$
\begin{equation}\label{eq:s1c}
\partial_\nu s^{\nu\lambda}=0.
\end{equation}
The resulting low-energy equations for $m_s\ll m_h$ are
therefore Eqs.~(\ref{eq:s2}) and (\ref{eq:s1c}),
without the need to impose a gauge condition.

Ultimately, this amounts to an effective simplification of
$\mathcal{L}_E+\mathcal{L}_m$ to
\begin{equation}
  \mathcal{L}_E+\mathcal{L}_m\to
  -\,\frac{1}{2\kappa}\left(\partial_\mu s_{\nu\lambda}\cdot\partial^\mu s^{\nu\lambda}
  +m_s^2 s_{\nu\lambda} s^{\nu\lambda} \right),
\end{equation}
keeping in mind that the tensor is constrained by Eq.~(\ref{eq:s1c}).

The standard gravitational coupling constant is the inverse
of the reduced Planck mass squared,
$\kappa=m_P^{-2}$. However, this can differ for towers of massive tensor states
or Kaluza-Klein states in large-volume compactifications, and therefore
we set $\kappa=f_s^{-2}$ and absorb $f_s$ into $s_{\nu\lambda}$ for canonical mass
dimension 1. $f_s^{-1}$ then appears as a coupling constant in all leading order
couplings of $s_{\nu\lambda}$.

\section*{Acknowledgments}

We acknowledge support from 
the Natural Sciences and Engineering Research Council of Canada.



\end{document}